# The Leeway of Shipping Containers at Different Immersion Levels[1]

Øyvind Breivik[*1,2], Arthur A Allen[3], Christophe Maisondieu[4], Jens-Christian Roth[5], Bertrand Forest[4]
2012-01-03

* Corresponding author. E-mail oyvind.breivik@met.no.
1 Norwegian Meteorological Institute, Alleg 70, NO-5007 Bergen, Norway
2 Geophysical Institute, University of Bergen, Norway
3 US Coast Guard Office of Search and Rescue, New London, CT, USA
4 IFREMER, Bretagne, France
5 Royal Norwegian Navy, Bergen, Norway

## Abstract

The leeway of 20-foot containers in typical distress conditions is established through field experiments in a Norwegian fjord and in open-ocean conditions off the coast of France with wind speed ranging from calm to 14 m s$^{-1}$. The experimental setup is described in detail and certain recommendations given for experiments on objects of this size. The results are compared with the leeway of a scaled-down container before the full set of measured leeway characteristics are compared with a semi-analytical model of immersed containers. Our results are broadly consistent with the semi-analytical model, but the model is found to be sensitive to choice of drag coefficient and makes no estimate of the cross-wind leeway of containers. We extend the results from the semi-analytical immersion model by extrapolating the observed leeway divergence and estimates of the experimental uncertainty to various realistic immersion levels. The sensitivity of these leeway estimates at different immersion levels are tested using a stochastic trajectory model. Search areas are found to be sensitive to the exact immersion levels, the choice of drag coefficient and somewhat less sensitive to the inclusion of leeway divergence. We further compare the search areas thus found with a range of trajectories estimated using the semi-analytical model with only perturbations to the immersion level. We find that the search areas calculated without estimates of crosswind leeway and its uncertainty will grossly underestimate the rate of expansion of the search areas. We recommend that stochastic trajectory models of container drift should account for these uncertainties by generating search areas for different immersion levels and with the uncertainties in crosswind and downwind leeway reported from our field experiments.

## 1 Introduction

Lost shipping containers represent a major safety risk for ship traffic as they can drift partly submerged for days before sinking. Also, the contents of shipping containers may be hazardous and thus represent a risk to health and environment (Breivik *et al*, 2011a, hereafter referred to as BAMR). The effort that goes into searching for containers or their potentially hazardous contents is the domain of trajectory models of drifting objects (Hackett *et al*, 2006; Breivik and Allen, 2008; Davidson *et al*, 2009) and sometimes oil spill models (Hackett *et al*, 2006). Successful backtracking of search objects to their origin using an iterative method with a stochastic trajectory model has also been demonstrated (Breivik *et al*, 2011b). The quality and precision of these forecast (and hindcast) services and related efforts like dynamical risk assessment of the danger of drifting objects (see e.g. Eide *et al*, 2007) is directly related to the precision of the drift properties that enter the trajectory estimates. This means that apart from all the usual uncertainties that arise from imperfect ocean current forecasts (clearly a large error source) or 10 meter wind forecasts, the uncertainty about the behavior of the object being searched for will have a major influence on the magnitude of the search area (Breivik and Allen, 2008; BAMR). Shipping containers are difficult to study in the field because of their size and weight and also because they represent a significant risk to their immediate surroundings if lost. Thus, despite their unarguable importance on the list of drifting objects, they have so far received little attention. The most important field study on shipping containers to date was presented by Daniel *et al* (2002), hereafter referred to as DJCLL. They studied the drift properties of 20-foot containers at various immersion levels and related their drift velocity to their immersion ratio using a semi-analytical relation. Unsurprisingly, the immersion level was found to be very important for the drift properties of containers. As immersion may vary significantly with loading and type of container (refrigerated containers will for example be more air-tight than regular containers and will thus remain afloat for a long time) it is important to assess the drift properties over the typical range of immersion levels.

---

[1] Revised version accepted for publication in *Ocean Dynamics, special issue on Advances in Search and Rescue at Sea*, DOI: 10.1007/s10236-012-0522-z

Trajectory models for estimating search areas for search and rescue (SAR) objects and hazardous material (HAZMAT) require estimates of the wind and wave induced drift in addition to the advection by the near-surface current (Hackett *et al*, 2006; Breivik and Allen, 2008; Davidson *et al*, 2009; BAMR). A review of earlier SAR and HAZMAT objects that have been the subject of field experiments was presented by Allen and Plourde (1999). They made recommendations for 63 categories of objects to be included in SAR planning tools. BAMR outlined a recommended field method for studying the drift properties of specific objects and provided worked examples on three typical SAR and HAZMAT objects, including a scaled-down (1:3.3 size) 40-ft container. Among their recommendations was the application of the following operational definition for the *Leeway* of a drifting object:

> Leeway is the motion of the object induced by wind (10 m reference height) and waves relative to the ambient current (between 0.3 and 1.0 m depth).

This definition standardizes the reference levels for the measurements of leeway for SAR objects and provides a practical way to utilize current and wind vectors from numerical models or measurements (e.g. high-frequency radars). The method outlined by BAMR can also be used to study relatively large objects, such as shipping containers, under certain assumptions.

This paper describes a procedure for performing leeway studies on shipping containers and compares the drift properties of a previously studied 20-ft container (DJCLL) and a scaled-down container (BAMR) with recent field work on 20-ft containers (described here). The paper is organized as follows. Section 2 describes the methodology followed and the field work conducted to collect leeway measurements on 20-ft containers along witch the leeway coefficients thus derived. Section 3 compares the drift properties from the field work with those derived from experiments on the earlier, scaled down container described by BAMR as well as the semi-empirical relation of DJCLL. A method for extrapolating the leeway coefficients derived for the specific immersion level is suggested based on DJCLL. The sensitivity of search areas to variations in the leeway coefficients thus derived is investigated using a stochastic ensemble trajectory model. Finally, Section 4 discusses further studies that will need to be conducted to establish a more robust estimate of the drift properties of shipping containers.

## 2 Field measurements of the leeway of 20-ft shipping containers

Field measurements of the leeway of two different 20-ft shipping containers were conducted in the Iroise Sea off the coast of Brittany, France, September 2008, and in the Andfjord in northern Norway, September 2009 (preliminary results were presented by Maisondieu and Forest, 2008 and Allen *et al*, 2010). The geographical positions of all the field work described or listed here is found in Figure 1 and details regarding the dates and instrumentation are found in Table 1. Experiment immersion levels are listed in Table 2.

The position, wind and leeway (the motion of the object through the ambient water) was measured directly from the object. This is referred to as the *direct leeway method* by BAMR. The weight and size of a container means that the difficulties normally encountered with measuring the wind and the leeway (the motion of the object through the ambient water) is less of a concern than for smaller objects. However, deployment and recollection of the container may pose serious difficulties. The various loading conditions found for drifting containers poses a challenge when trying to establish typical leeway estimates. It seems clear that it is necessary to somehow follow DJCLL and relate the leeway to the immersion level.

Field experiments on the leeway of containers can in principle follow the outline described by BAMR, but care has to be taken when interpreting the reference depth of the ambient current as the object has a deeper draught than typical SAR objects (2 m is a typical draught found with containers, while life rafts will have a draught of maybe 30 cm).

### *2.1 Experiment design*

The nominal exterior dimensions of the 20-ft standard shipping container are 6,058 mm by 2,438 mm by 2,591 mm (height). The volume is 33.131 $m^3$ and it weighs 2,229 kg when empty.

Two similar experiments were conducted on 20-ft containers (Iroise Sea, Figures 2-3; Andfjord, Norway, Figures 4-5). In both cases holes were made to allow the container to fill to the desired immersion and drain when hoisted up after the experiment. The total area of the apertures was approximately 1.5 $m^2$, large enough to allow the container to rapidly fill and drain while keeping the water flow in and out of the container at an acceptable level. Buoyancy was provided by placing barrels under the ceiling (see Figures 2 and 4). Total additional weight of the containers differed slightly but was 400-450 kg (see details below). The containers were equipped with Automatic Identification System (AIS) transponders that allowed continuous monitoring from a nearby research vessel. This had the added advantage of making the object more visible to nearby vessels.

Records of wind, position and leeway were matched in time for analysis on 10-minute samples. The leeway was decomposed in components aligned with the downwind direction and orthogonal (crosswind) to the downwind direction for every 10-minute sample (see Allen, 2005; Breivik and Allen, 2008; BAMR). The wind was adjusted to 10-m height following Smith (1988).

**Experimental setup, Iroise Sea 2008**
Drift trials were conducted in September 2008 in the Iroise Sea off the coast of Brittany, France, with a regular 20-ft container (see Figure 1 and Table 1). The supply vessel BSAD Alcyon is chartered by the French authorities for counter-pollution operations and was adequately fitted for handling deployment and recovery of 20-ft containers. The container was equipped with 24 barrels, fixed under the ceiling, for buoyancy. Total additional buoyancy was 5.75 m$^3$; 440 kg of weights were added and evenly placed on the floor for stability. Apertures were made on both floor and ceiling in order to allow water filling when dropping the container. With such an arrangement, the container was set to float horizontally with a freeboard of 59 ± 5 cm (77±2 % immersion).

The leeway was measured by an acoustic current meter (Nortek Aquadopp) attached by a 20 m line. The measurement cell was centered approximately 3 m below the surface. Drag of the current meter was partially compensated by wind action on the supporting buoy and could be considered negligible compared to the container's inertia. Only horizontal components of the flow were recorded. Measurement sampling period was 5 s averaged over 5-minute intervals. Accuracy on flow velocity measurement is 1.1 cm s$^{-1}$. Local wind was measured by a sonic anemometer fixed to a rigid mast vertically mounted on top of the container at an elevation of 2.35 m±5 cm above the sea surface. In order to get the absolute wind direction, the anemometer was coupled with the compass of a Motion Record Unit onboard the container. Record sampling rate was 1 Hz. Wind measurements were averaged over 10-minute periods.

The duration of the drift trial was 5 hours (Table 1) during which the wind direction remained north-easterly with a speed around 13 m s$^{-1}$ (10 min averaged at 10 m, see Table 3). Sea state, as observed by a directional wave buoy located a few nautical miles to the east of the trial area was dominated by a rather long swell (average peak period of 13.1 s) traveling from west-northwest (average peak direction around 284°). Total significant wave height was around 0.8 m. Tidal currents in the trial area are mostly oriented south during ebb and north during flood and can reach rather high velocities, up to almost 2 m s$^{-1}$ during spring tides, but were constantly below 1 m s$^{-1}$ during the experiment.

**Experimental setup, Andfjord 2009**
The second field experiment was carried out with a similarly equipped container in Andfjord in northern Norway. Two drift runs were conducted in the Andfjord between 24 and 27 September 2009 by the Norwegian Coast Guard Vessel Harstad (Table 1). The Andfjord is a semi-enclosed basing with a wide mouth to the NW. The wind varied from 1 to 14 m s$^{-1}$ with moderate wind sea of short fetch and no swell intrusion into the fjord.

Barrels under the ceiling provided flotation. Holes were cut in the bottom and sides of the container to allow the container to fill and drain during deployment and recovery (see Figure 4). The total buoyancy was slightly less than for the Iroise Sea experiment and the total weight was also slightly less (ca 400 kg extra weight), giving a freeboard of 50±5 cm, i.e. an immersion ratio of 80±2%; slightly deeper than in the Iroise Sea experiment. The container had a sonic anemometer and a Coastal Environmental Systems WeatherPak wind monitoring system mounted on the roof (see Figure 5). The leeway was measured using an upward looking Nortek AquaDopp acoustic Doppler current profiler (ADCP) in tow behind the container as well as an InterOcean S4 Electromagnetic current meter (EMCM) attached via a 30 m line.

## *2.2 Leeway coefficients*

The leeway speed, downwind component of leeway (DWL) and the crosswind components of leeway (CWL) 10-minute data for the four drift runs of the container versus the wind speed adjusted to the 10-m height are shown in Figures 6-8 along with the unconstrained and constrained linear regressions and their respective 95% prediction limits following the procedure outlined by BAMR. We calculated the leeway slope and intercept of an unconstrained linear regression for both leeway speed *v* wind speed, DWL and the CWL, respectively, following BAMR,

$$L = aW_{10} + b + \varepsilon, \quad (1a)$$
$$L_d = a_d W_{10} + b_d + \varepsilon_d. \quad (1b)$$
$$L_c = a_c W_{10} + b_c + \varepsilon_c, \quad (1c)$$

Here $W_{10}$ is the 10 m wind speed, *a* (regression slope) is denoted the leeway rate (leeway to wind ratio), *b* is the regression intercept at zero wind and $\varepsilon$ the regression residual. Eq (1a) relates the leeway speed to the 10-m wind speed while Eqs (1b)-(1c) give similar estimates for the vector components of the leeway in the downwind (subscript d) and crosswind (subscript c) directions.

The leeway speed and the downwind components were regressed against the 10-m wind speed. The linear regression was done both unconstrained and constrained through the origin for each of the two experiments described here (see Tables 3-4). We have also listed the results previously reported by BAMR describing a scaled-down size 1:3.3 40-ft container. The aggregated leeway data for all three field experiments are shown in Figures 6-8. The combined leeway coefficients are listed in Tables 3 and 4. The leeway rate *a* (regression slope), the intercept *b* (for the unconstrained regression), the standard error for the regression estimate, $S_{yx}$, and $r^2$ were calculated following Neter *et al* (1996). Figures 6-8 show the constrained and unconstrained regression lines with their associated 95 % confidence interval, corresponding to $\pm 2 S_{yx}$. For the crosswind components of leeway the values were separated along runs or portions of runs indicated by the progressive vector diagrams to be consistently left (negative) or right (positive) of the downwind direction (see Section 2.3 on jibing below). The positive crosswind estimates were combined with the negative crosswind estimates by changing the sign of the latter before regressing against the 10-m wind speed. Again both unconstrained and constrained estimates were made (see Tables 3-4).

Allowing for a regression intercept brings down the standard error of the regression estimate. The reason for a zero offset can be instrument or experiment error, but may also indicate the presence of wave effects not accounted for at low winds. We do not study wave effects explicitly here but follow Breivik and Allen (2008) and argue that in high wind situations the Stokes drift is directly related to the wind speed and thus contained in the leeway rate. This assumption breaks down for low-wind conditions in the presence of swell or in situations with rapidly changing wind direction, such as front passages.

### *2.3 Jibing*

Allen (2005) introduced the concept of jibing of drifting objects. Jibing refers to an abrupt change of the (stable) drift direction relative to the downwind direction. Jibing is observed in leeway drifts as significant and prolonged sign changes of the crosswind component of leeway (CWL). The jibing can be considered a binomial process, i.e., within a certain time period the object either changes its orientation relative to the wind or it does not. By the central limit theorem (see e.g. Press *et* al, 1992), a stochastic trajectory model which takes into account jibing will tend towards a stronger central tendency. The frequency of these changes in CWL is an important factor when modeling search areas as an object with a high jibing frequency will tend to "fill in" the middle part of a search domain between the two high-probability regions of persistently right-drifting and left-drifting objects (Breivik and Allen, 2008).

Jibing or changes in CWL sign occur either abruptly from one ten-minute sample to the next or gradually over several sampling periods where the CWL value remains within our ability to distinguish CWL from zero. The second method requires subjective inspection of the time series to identify negative (left-of-downwind) or positive (right-of-downwind) segments. A progressive vector diagram of the leeway of the object is used to identify the jibing events (indicated by an arrow in Figure 9). Allen (2005) provided an estimate of the jibing frequency of 3 and 7 % per hour based on the available leeway experiments to date. We observed only one jibing event during a total of more than 40 hours, roughly consistent with a jibing frequency of 3-7 %.

## 3 The variation of leeway with immersion

It is known that the slip or leeway of an object will be a function of the level of immersion (Kirwan *et al,* 1975; Niiler *et al* 1987; Geyer, 1989; O'Donnell *et al,* 1997),

$$I = \frac{A_w}{A_a + A_w}, \quad (2)$$

where *I* is the immersion ratio and $A_a$ and $A_w$ the area of the over-water and under-water cross sections.

The same functional relationship has also been applied to icebergs (Smith, 1993). DJCLL fitted experimental data to an analytical model for the leeway rate as a function of the immersion level of containers. Following DJCLL, Eqs (12-13), with a slight change of notation and retaining the drag coefficients of air and water as free parameters, we get the following relation between the leeway rate and the immersion level (Eq 2),

$$\hat{a} = \frac{L}{W_{10}} = \frac{1-I-\sqrt{\frac{\rho_a C_a}{\rho_w C_w}\frac{1-I}{I}}}{1-(1+\frac{\rho_a C_a}{\rho_w C_w})I}, \quad (3)$$

where $L$ is the leeway speed, $W_{10}$ the 10-m wind speed and $\hat{a}$ is the modeled leeway rate. The density ($\rho$) and drag coefficient ($C$) of air and water are marked with suffixes "a" and "w" respectively. BAMR simplified the above expression to read

$$\hat{a} \approx \sqrt{\frac{\rho_a C_a}{\rho_w C_w}\frac{1-I}{I}}. \quad (4)$$

Eq (4) will for immersion levels above 50 % deviate by less than 5 % from Eq (3). Figure 10 compares the leeway rate $\hat{a}$ computed using constrained linear regression found in Table 3 at the measured immersion levels with Eq (3). The leeway rate increases very quickly with immersion for levels below 40 %, but for deeper (and more realistic) draughts the dependence on immersion is weaker and almost linear. The 95 % confidence level for the observed immersion level is shown by a horizontal bar for each data set (the observed immersion levels are listed in Table 2). The uncertainty in leeway rate is harder to estimate. We follow Breivik and Allen (2008) and note that since the spread about the regression line is slightly heteroscedastic (increasing observational spread with increasing wind speed), the regression *slope* (leeway rate) should be perturbed to account for the spread over the typical range of wind speed from 0 to 20 m s$^{-1}$. It is then not enough to calculate the standard error of the regression slope as this assumes homoscedastic data. The spread should be comparable to the observed spread (see the red dashed line in Figure 6 and the estimates in Table 3) at the mid point of this wind speed range, i.e. at 10 m s$^{-1}$ (see Breivik and Allen, Eqs 6-7). It follows that the uncertainty in the leeway *rate* is $a \pm S_{yx}/10$ (the vertical bar in Figure 10). DJCLL assumed that the drag coefficient for water would lie between 0.8 and 1.2 for containers following experimental work by Cabioc'h and Aoustin (1997) and chose 1.0 as an average water drag coefficient. They also set the air drag coefficient to 1.0 arguing that this would be a reasonable value for bluff objects at high Reynolds numbers. To investigate the dependence on the air drag coefficient we also calculated corresponding immersion curves for coefficients of 0.7 and 1.15. This range encompasses the constrained leeway rates estimated for the three containers (see Table 3) shown in Figure 10. Our observed leeway rate appears broadly consistent with the results found by DJCLL for the three containers studied here at $C_a$=1.0. It is evident from Figure 10 that the leeway rate is very sensitive to the immersion ratio at low immersion levels but only moderately sensitive in the range from 50 % immersion. It is also evident that the leeway rate is sensitive to the drag coefficient. This wide range of leeway rates resulting from moderate changes in the drag coefficient and immersion level is an obvious source of uncertainty when applying the immersion model of DJCLL. On the other hand, immersion levels below 50 % must probably be handled differently as the container would become unstable and start to roll. This may in fact happen, in particular for empty refrigerated containers, but these will not be considered here. We argue that it may reasonably be assumed that most containers will have an immersion level higher than 50 % and will in the following focus on the range 60-90 %.

As BAMR pointed out, another weakness of the semi-analytical model of DJCLL is that no estimate is made of the crosswind leeway component. Without this, the later extent (perpendicular to the general drift direction) of search areas may be underestimated, even if the leeway rate is roughly correct. The leeway divergence can be quite substantial (cf Figures 8-9), as is also evident from the relatively high crosswind leeway coefficients listed in Tables 3 and 4. The leeway divergence angle is the average angle between the wind direction and the drift direction of the object (Allen, 2005; Breivik and Allen, 2008). It may be found from the crosswind and downwind leeway rates,

$$\bar{\alpha} = \tan^{-1}\left(\frac{a_c}{a_d}\right) \quad (5).$$

The average leeway divergence angle of containers estimated from our field measurements is 17º (using the constrained combined values in Table 3 for $a_d$ and $a_c$), but it is evident from Figure 9 that the leeway divergence varies significantly. There is the possibility that the Iroise Sea experiment suffered from a compass error (red curve in Figure 9). However, to avoid underestimating the leeway divergence we have chosen to retain it. Note also that the leeway *speed* estimates remain unaffected by compass error. When searching for the object, a divergence angle of 17º translates into a total angle of uncertainty twice that (34º) since we have no way of telling in advance whether an object will orient itself to the right or to the left of the general wind direction. Consequently, we must assign equal probability to both outcomes and compute trajectories that veer off to the left and to the right of the downwind direction. The leeway divergence angle may well vary with immersion level, but lacking measurements at other immersion levels we keep it constant. Indeed, there is considerable measurement uncertainty associated with the leeway divergence angle. By using the unconstrained coefficients in Table 4 the divergence angle increases to 24º. However, we have for consistency computed the leeway divergence angle from the constrained coefficients as we use these in the following analysis. By scaling the crosswind and downwind coefficients listed in Table 3 with the leeway rate in Eq (3), we have compiled leeway coefficients for the immersion range 50-90 % in Table 5.

To investigate the sensitivity of the rate of expansion of search areas to immersion level, air drag coefficient and crosswind component (or leeway divergence), we performed a series of idealized simulations with the *Leeway* model, a stochastic trajectory model (Breivik and Allen, 2008). A constant southerly wind 10 m s$^{-1}$ and no current was applied in an open ocean location with no risk of stranding particles. Ensembles of 500 members were released at immersion levels of 60 % and 90 %, bounding the reasonable immersion range for containers. All ensemble members were released from a single point (release positions are marked with "*" in blue in Figure 11. For clarity of presentation the release positions of the ensemble integrations of the two immersion levels were separated horizontally. The search areas after 48 hours from release are shown as convex hulls enclosing the individual ensemble members (not shown) in Figure 11. Three drag coefficients were tested, 0.7 (red), 1.0 (black) and 1.15 (red dashed). Since we are uncertain about the actual leeway divergence angle at other immersion levels it is important to assess the impact of crosswind measurement uncertainty on the rate of expansion of the search areas. Therefore, a leeway divergence angle of 17º, computed using Eq (5), was used for the high and low drag coefficients (red polygons), while the middle (air drag $C_a$=1.0) assumed no crosswind but kept the same crosswind standard error ($S_{yx}$=4.86) found in Table 3. The stochastic trajectory model perturbs the ensemble according to the standard error of the regression estimate in crosswind and downwind coefficients ($S_{yx}$, see Table 3), but uses the immersion-corrected leeway rate listed in Table 5. This allows a comparison of the importance of the various sources of uncertainty contributing to the evolution of a container search area; (i) the rate of expansion of the search, (ii) the errors in drift coefficients (uncertainties in air drag coefficients and lack of crosswind leeway and (iii) the level of immersion of the container. As Figure 11 shows, the most important error source is the level of immersion since making a wrong guess here might potentially more than halve the leeway of the object (easily seen by comparing the leeway rate at 60 % to that at 90 % in Figure 10). However, the choice of air drag coefficient is also important. Reducing the air drag to 0.7 reduces the speed of the object by about 25 %. The choice of a more realistic leeway divergence angle expands the search area at 60 % immersion by approximately 25 % compared to setting the crosswind to zero (black polygon). Note that because we use the crosswind standard error found experimentally, the rate of expansion of the search area for zero crosswind is much wider (and more realistic) than what would be found simply by using the original equation of DJCLL where the only coefficient that can be perturbed is the immersion level (or possibly the air drag coefficient). For reference we have computed the displacement according to DJCLL's original formulation (Eq 3) for immersion levels 60, 70, 80 and 90 %, with no error estimate in the crosswind or downwind leeway coefficients. Clearly, even for an ensemble with realistic ranges in the immersion level, the search area will be grossly underrepresented as no estimate of the leeway divergence and the crosswind expansion of the search area is made.

## 4. Discussion and Concluding Remarks

We have described an experimental method to collect leeway measurements on shipping containers. The data collected cover two 20-ft containers immersed to roughly 75-83 %. The results were aggregated with field studies of a scaled-down container (see BAMR) and compared with the semi-analytical immersion model developed by DJCLL.

The recommended procedure for leeway field measurements outlined by BAMR was followed in our field experiments. Two problems arise when studying objects of this size. First, the draught of the container in the typical distress configuration of around 80 % is 2 m. This means that measuring leeway at 50 cm depth may no longer be the best level. However, as the leeway estimates will eventually serve as input to trajectory models which rely on near-surface current estimates from either numerical models or observations (Breivik and Allen, 2008), it is important to relate the leeway coefficients to the same reference level as for smaller SAR objects. Secondly, the wave effects are ignored in our studies as they have been in earlier leeway field studies. The argument that wave effects (wave excitation and damping, see Mei, 1989) are negligible for small SAR objects (Hodgins and Hodgins, 1998; Breivik and Allen, 2008) may not be strictly true for an object the size of a 20-ft container. However, we do argue that for objects smaller than 30 m the wave effect and also the Stokes drift (Mei, 1989; Holthuijsen, 2007) will be related to wind sea rather than swell. Hence, the *direction* of the wave-induced drift will broadly coincide with the direct wind effect on the over-water structure of the container and the leeway coefficients estimated using the direct method (Eqs 1 a-c) will include the wave effect.

We found that the results agree fairly well with those obtained by DJCLL. However, small variations in air drag coefficients and immersion level will modify the leeway rates significantly (see Figure 10). A more serious weakness of the model DJCLL is that crosswind leeway is ignored completely, i.e., no leeway divergence is expected and the object will essentially drift downwind. This will lead to errors in the drift direction but more importantly by ignoring the experimental *uncertainty* in the crosswind leeway the search area will not exhibit the required lateral expansion to properly account for the true uncertainty of the search (see Figure 11). It is clear that in order to properly assess the validity of the semi-analytical model of DJCLL, leeway measurements will have to be collected at the range of typical immersion ratios (60-90%). However, it appears that a reasonable estimate of the leeway rate can be found using extrapolation of the empirical crosswind and downwind leeway coefficients estimated from the field experiments described in Section 2 by using Eq (3) with the leeway divergence assumed constant over the immersion range (see Table 5 and Figure 11).

We recommend that the drift of containers be modeled by extending the immersion ratio model of DJCLL to include the cross wind coefficients found empirically here (see Table 5) and with perturbations to downwind and crosswind estimates based on the combined $S_{yx}$ error estimates listed in Tables 3-4. The air drag coefficient should be in the range 0.7 to 1.15. In light of the high sensitivity of a container's leeway to immersion, we further recommend that stochastic trajectory models take this uncertainty into account by generating ensembles based on a reasonable range of immersion levels (60-90 %).

Acknowledgments
The authors would like to acknowledge the substantial support that this work has received through the ship time allotted by the Norwegian Coast Guard and the French Navy, "Action de l'Etat en mer" without which there would be no field work. The French field work was partly funded by the LostCont Interreg project, where CEDRE and SASEMAR played a vital role in facilitating the field studies. The Norwegian field work received funding from the Norwegian Navy and the FOB project funded through the Research Council of Norway MAROFF programme (grant no. 180175). The analysis has benefited from the MAROFF projects FARGE (grant no. 200843) and "Uncontrolled drift of ships and larger objects" (grant no. 200862). The work built on results from the French-Norwegian SAR-DRIFT project (Eureka E!3652). The authors also wish to thank the Norwegian Joint Rescue Co-ordination Centres (JRCC) and the Norwegian Navy for their continued support of the development of operational trajectory models for search and rescue.

| Run | Start time (UTC) | End Time (UTC) | Duration (hh:mm) | Current meter sampling depth (m) | Wind measurement height (m) | 10-m wind speed range (m/s) |
|---|---|---|---|---|---|---|
| Iroise Sea | 2008-09-22 13:00 | 2008-09-22 18:00 | 5:00 | 2 MHz Aquadopp (1.5 – 2.2) | CV7 Sonic (2.35 m) | 9.6-13.0 |
| Andfjord Run 1 | 2009-09-24 09:10 | 2009-09-25 05:20 | 8:40 | S4 (0.75) & 1 MHz Aquadopp (0.60 -1.12) | CV7 Sonic (2.5 m) | 1.3-10.2 |
| Andfjord Run 2 | 2009-09-27 05:40 | 2009-09-27 / 16:40 | 11:00 | S4 (0.75) & 1 MHz Aquadopp (0.60 -1.12) | CV7 Sonic (2.5 m) & R.M Young w/Weatherpak (2.5 m) | 4.3-14.3 |

Table 1. 20-ft Container experiment overview.

| Drift Object | Freeboard (cm) | Container height (cm) | Immersion (%) |
|---|---|---|---|
| Model 40-ft | 25 +/-3 | 79 | 68.3 +/- 3.9 |
| Full sized 20-ft Andfjord | 50 +/- 5 | 259.1 | 80.7 +/- 2.0 |
| Full sized 20-ft Iroise Sea | 59 +/- 5 | 259.1 | 77.2 +/- 2.0 |

Table 2. Freeboard and immersion levels of the drift objects.

| Drift Object | 10-m wind speed (m s$^{-1}$) | 10-min samples | Leeway Speed | | | DWL | | | CWL | | |
|---|---|---|---|---|---|---|---|---|---|---|---|
| | | | $a$ (%) | $S_{yx}$ (cm s$^{-1}$) | $r^2$ | $a_d$ (%) | $S_{yx}$ (cm s$^{-1}$) | $r^2$ | $a_c$ (%) | $S_{yx}$ (cm s$^{-1}$) | $r^2$ |
| Modeled 40-ft | 0.1 – 11.1 | 95 | 2.00 | 3.03 | 0.71 | 1.96 | 3.08 | 0.83 | 0.02 | 2.65 | -0.11 |
| Full 20-ft Iroise Sea | 9.6 – 13.0 | 30 | 1.71 | 1.34 | 0.25 | 1.40 | 2.69 | 0.05 | 0.93 | 3.32 | 0.02 |
| Full 20-ft Andfjord | 1.3 – 14.3 | 118 | 1.83 | 3.83 | 0.39 | 1.73 | 3.22 | 0.53 | 0.33 | 4.36 | 0.01 |
| Combined | 0.1 – 14.3 | 243 | 1.82 | 3.28 | 0.65 | 1.64 | 3.62 | 0.61 | 0.51 | 4.83 | 0.17 |

Table 3. Linear Regression Parameters constrained through zero for leeway speed, downwind component of leeway (DWL) and crosswind component of leeway (CWL). Slope of regression line (leeway rate) is denoted $a$, standard error $S_{yx}$ for the regression and correlation squared $r^2$.

| Drift Object | Leeway Speed | | | | DWL | | | | CWL | | | |
|---|---|---|---|---|---|---|---|---|---|---|---|---|
| | $a$ (%) | $b$ (cm s$^{-1}$) | $S_{yx}$ (cm s$^{-1}$) | $r^2$ | $a_d$ (%) | $b_d$ (cm s$^{-1}$) | $S_{yx}$ (cm s$^{-1}$) | $r^2$ | $a_c$ (%) | $b_c$ (cm s$^{-1}$) | $S_{yx}$ (cm s$^{-1}$) | $r^2$ |
| Modeled 40-ft | 1.426 | 4.757 | 3.029 | 0.921 | 1.786 | 1.44 | 2.995 | 0.839 | -0.27 | 2.44 | 2.31 | 0.17 |
| Full 20-ft Iroise Sea | 1.167 | 6.481 | 1.285 | 0.320 | 0.916 | 5.76 | 2.70 | 0.062 | 0.70 | 2.67 | 3.35 | 0.02 |
| Full 20-ft Andfjord | 1.22 | 5.08 | 3.25 | 0.57 | 1.25 | 3.96 | 2.81 | 0.65 | 0.19 | 1.14 | 4.36 | 0.02 |
| Combined | 1.30 | 4.91 | 2.46 | 0.80 | 1.27 | 3.58 | 3.26 | 0.685 | 0.59 | -0.72 | 4.82 | 0.18 |

Table 4. Unconstrained Linear Regression Parameters. Slope of regression line (leeway rate) is denoted $a$, offset $b$, standard error $S_{yx}$ for the regression and correlation squared $r^2$.

| Immersion level | 50% | 55% | 60% | 65% | 70% | 75% | 80% | 85% | 90% |
|---|---|---|---|---|---|---|---|---|---|
| CWL avg ($C_a$=1.0) | 0.96 | 0.87 | 0.79 | 0.71 | 0.64 | 0.56 | 0.49 | 0.41 | 0.33 |
| DWL avg | 3.09 | 2.80 | 2.54 | 2.29 | 2.05 | 1.81 | 1.57 | 1.33 | 1.05 |
| CWL high ($C_a$=1.15) | 1.03 | 0.93 | 0.85 | 0.76 | 0.68 | 0.60 | 0.52 | 0.44 | 0.35 |
| DWL high | 3.31 | 3.00 | 2.72 | 2.45 | 2.19 | 1.94 | 1.68 | 1.42 | 1.13 |
| CWL low ($C_a$=0.7) | 0.81 | 0.73 | 0.66 | 0.60 | 0.53 | 0.47 | 0.41 | 0.35 | 0.27 |
| DWL low | 2.60 | 2.36 | 2.14 | 1.92 | 1.72 | 1.52 | 1.32 | 1.11 | 0.88 |

Table 5. Estimated DWL and CWL rates (%) for 50 to 90 % immersion. The average (avg) values correspond to an air drag coefficient $C_a$=1.0; upper estimates correspond to 1.15 and lower estimates to 0.7 (see Figure 10). The leeway divergence angle is assumed constant at 17º. Regression standard error for DWL and CWL

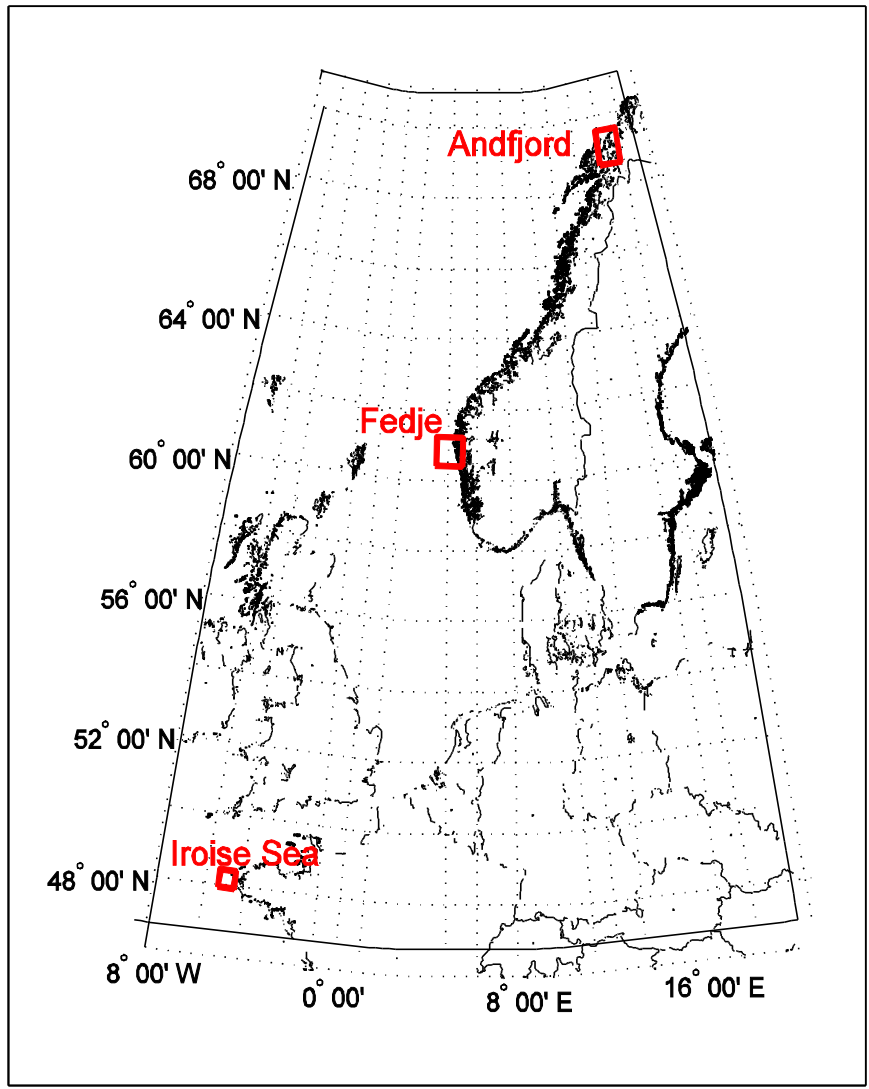

Figure 1. Experiment areas Fedje and Andfjord, Norway, and the Iroise Sea, France.

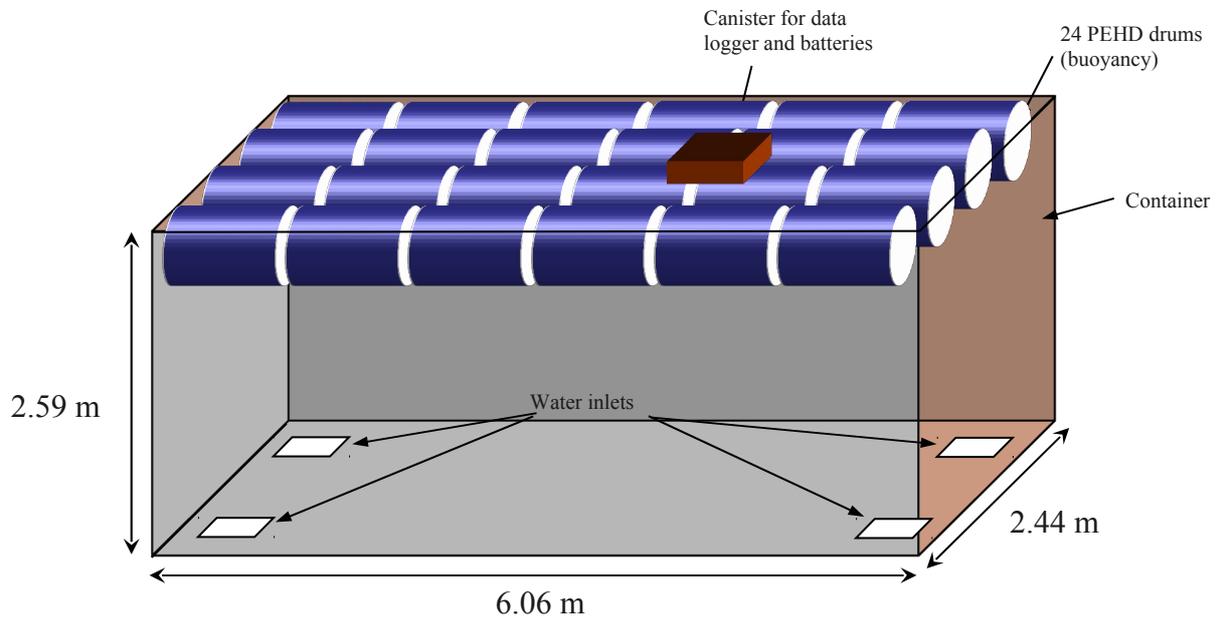

Figure 2. The Iroise Sea experiment. Sketch of container with buoyancy and water inlets. The average immersion level was 77% with a total weight of 2897 kg.

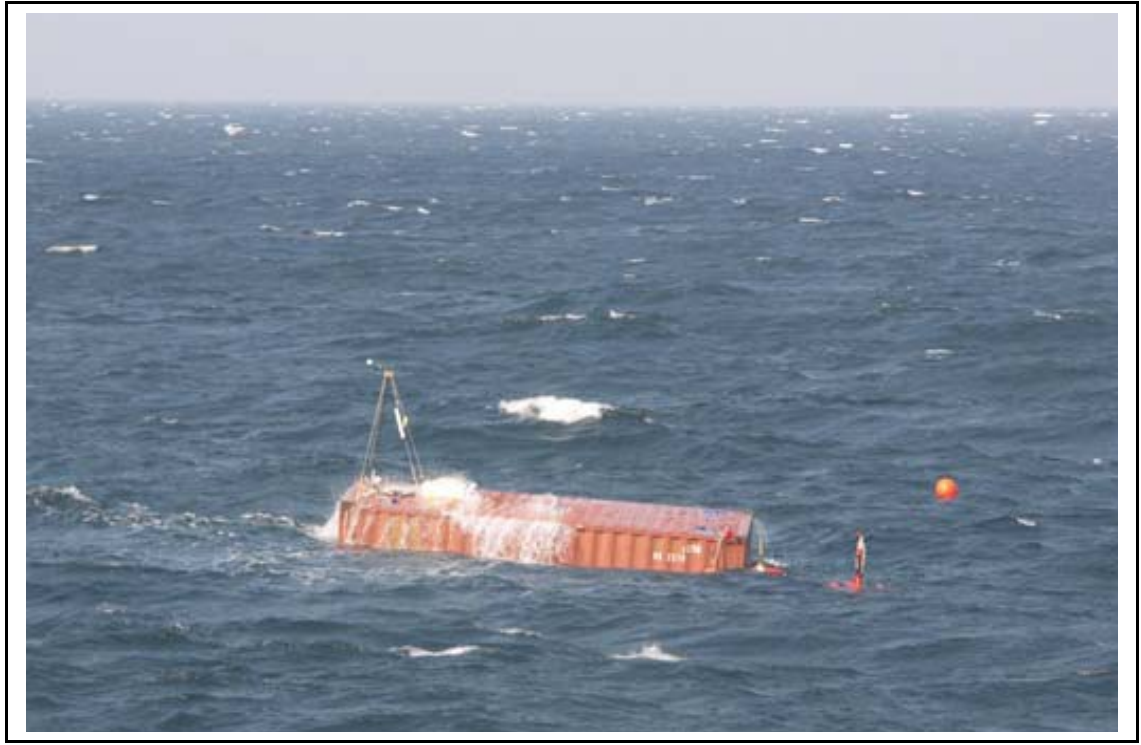

Figure 3. The Iroise Sea experiment. The 20-ft container deployed in the Iroise Sea, showing the sonic anemometer (left side of container) and the red float with the upward looking current meter suspended from it.

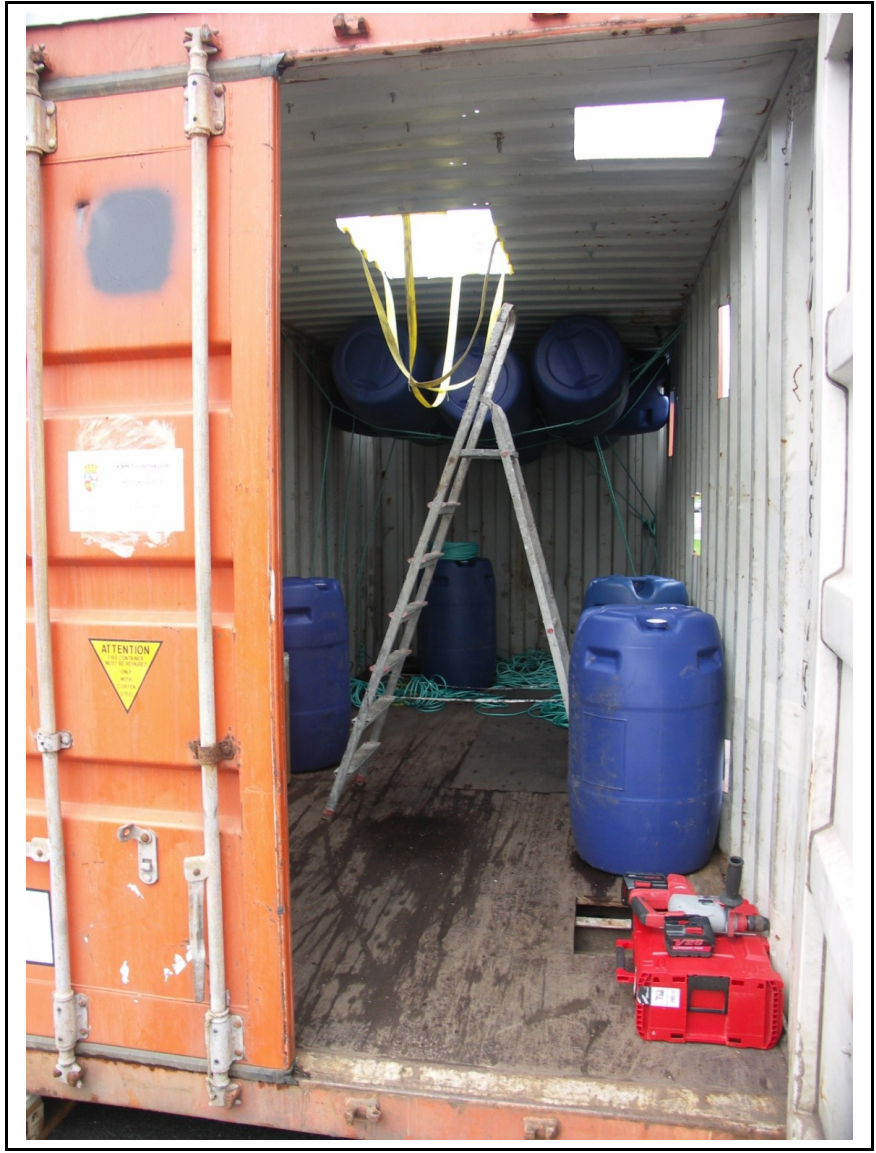

Figure 4. The Andfjord experiment. Inside of container, showing some of the plastic barrels hung from the ceiling of the container to provide buoyancy. The sling in the middle of ceiling is to hold the sonic anemometer electronics and batteries, and the AIS beacon. Holes were made in the walls as well as the floor for rapid immersion.

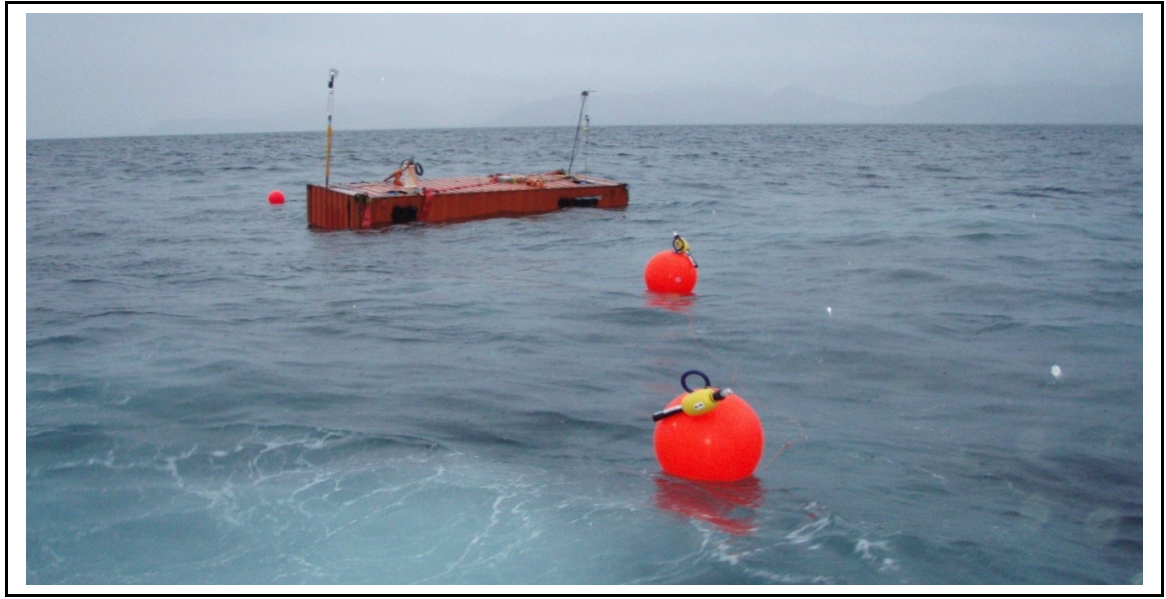

Figure 5. The Andfjord experiment. The 20-ft container is deployed with a WeatherPak system on the left and sonic anemometer on the right. The two orange floats support the S4 and Aquadopp current meters.

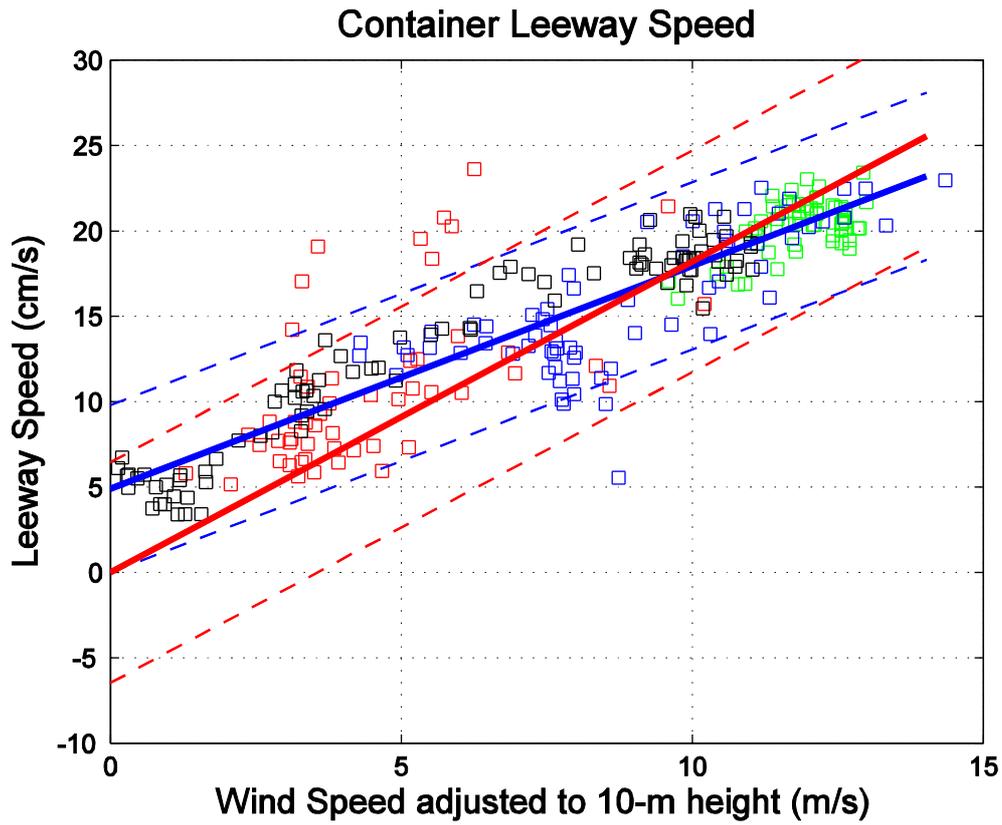

Figure 6. Leeway Speed *v* wind speed adjusted to 10-meter height for the four container drift runs. Green squares are from the Iroise Sea (20ft), red and blue from the Andfjord (20ft), and black from the scaled-down container. Unconstrained linear regression mean (solid) and 95% confidence levels (dashed) corresponding to ±2$S_{yx}$ are plotted in blue. Linear regression constrained through zero and its associated 95% confidence intervals (dashed) are shown in red.

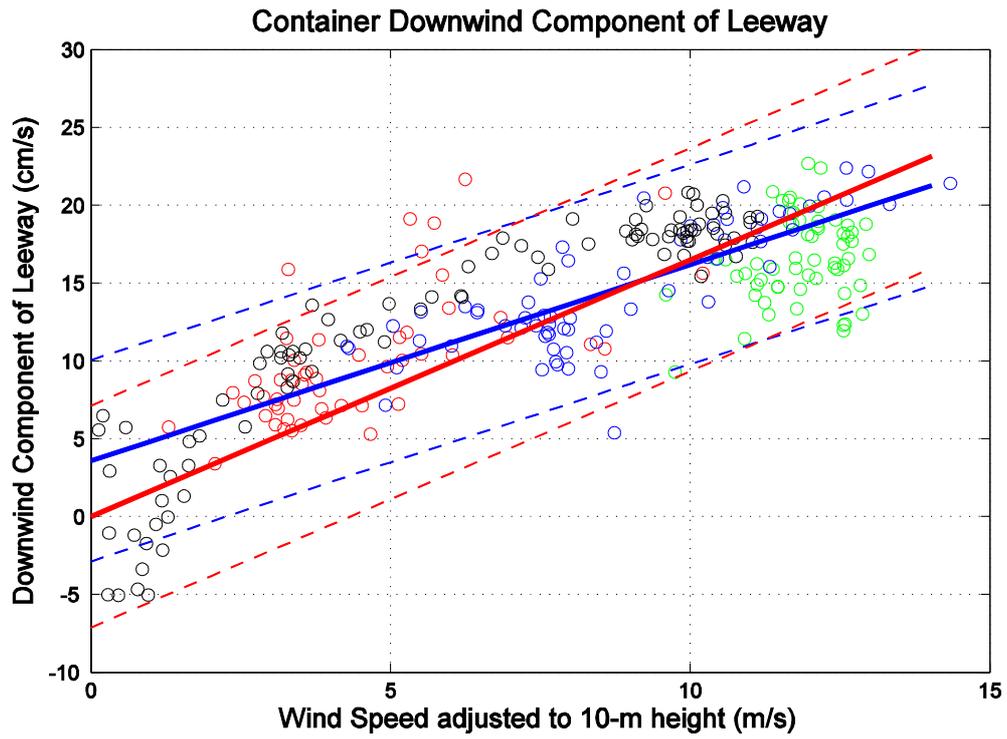

Figure 7. Downwind Component of Leeway (DWL) *v* wind speed adjusted to 10-m height for the four container drift runs (colors same as for Figure 6).

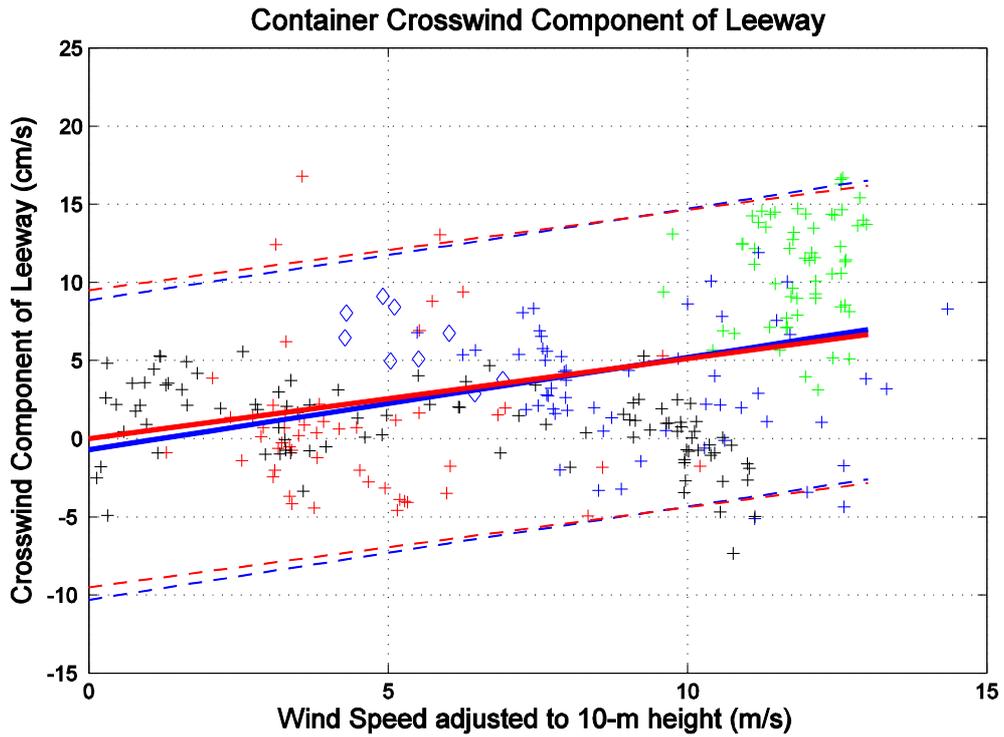

Figure 8. Crosswind Component of Leeway (CWL) *v* wind speed adjusted to 10-m height for the four container drift runs (colors same as for Figure 6). Blue diamonds indicate left-drifting leeway estimates following a jibe in the Andfjord experiment.

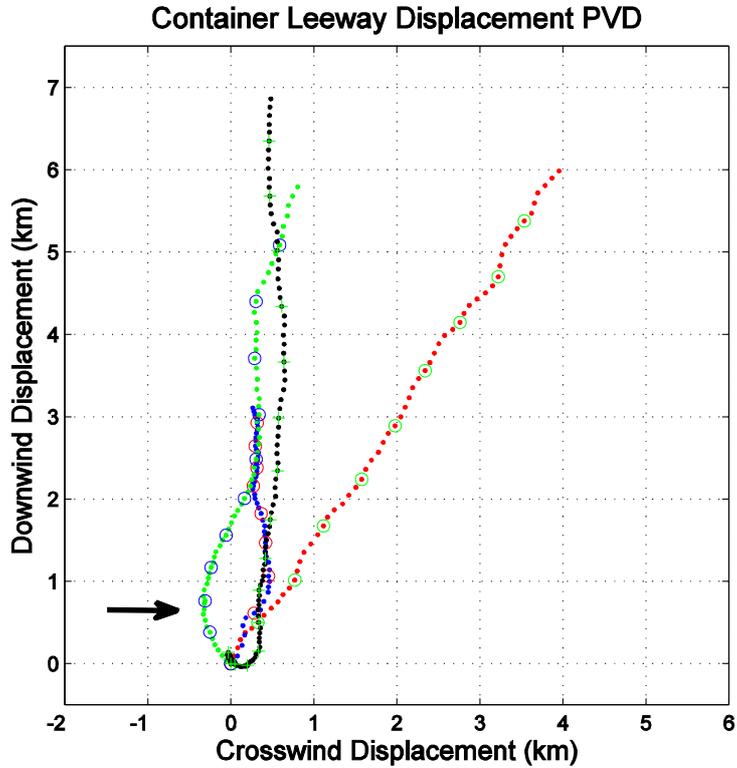

Figure 9. Progressive vector diagram (PVD) of the downwind and crosswind components of leeway displacement for the four shipping container drift runs. Red is Iroise Sea, blue and green are Andfjord runs 1 and 2, and black is modeled container from Fedje. Downwind is up, i.e., an object drifting directly with the wind will show up as a trajectory pointing upwards. Positive crosswind is to the right. Displacement is in kilometers. Hourly intervals are indicated by a circle every 6$^{th}$ point. The average leeway divergence estimated from Eq (5) is 17º, but it is evident that the leeway divergence varies significantly. One jibing event was identified by visual inspection (indicated by the black arrow). The large divergence angle found in the Iroise Sea experiment (red curve) may be due to a compass error. However, the leeway speed is confirmed to be consistent with the other experiments.

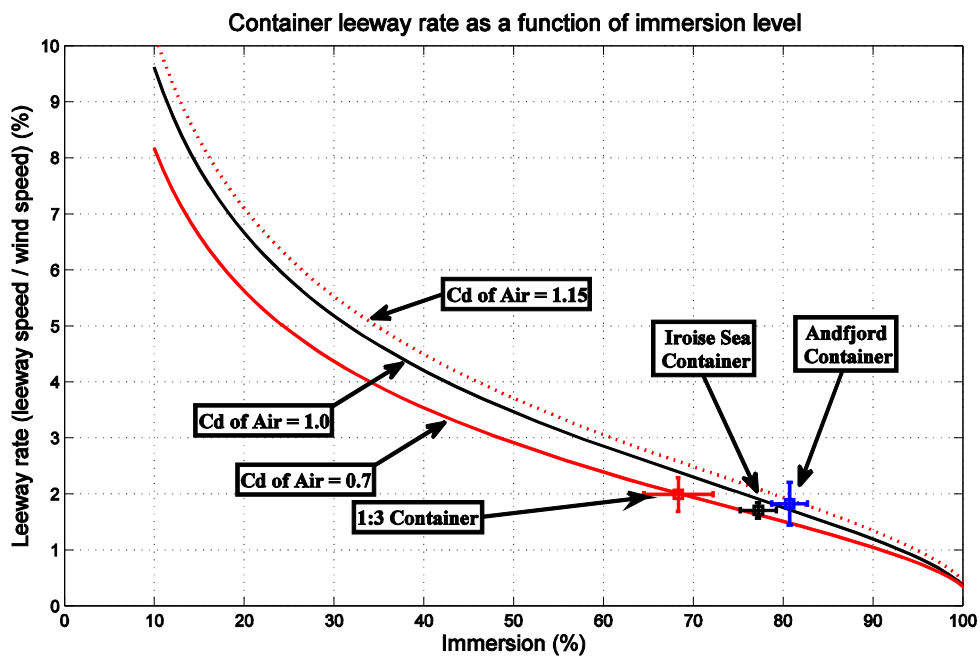

Figure 10. The leeway rate (leeway to wind speed ratio) as a function of immersion level (%) following Eq (3) [Eq (16) of Daniel *et al* (2002)]. The black line corresponds to an air drag coefficient of 1.0 (similar to Daniel *et al*, 2002). Upper and lower bounds are found by computing the immersion curve for a drag coefficient of 0.7 (red dashed line) and 1.15 (red full line). The three objects described and studied here are shown as crosses indicating the uncertainty in immersion and the leeway rate $a \pm S_{yx}/10$ from the (Iroise Sea black, Andfjord blue and scaled-down container marked as red).

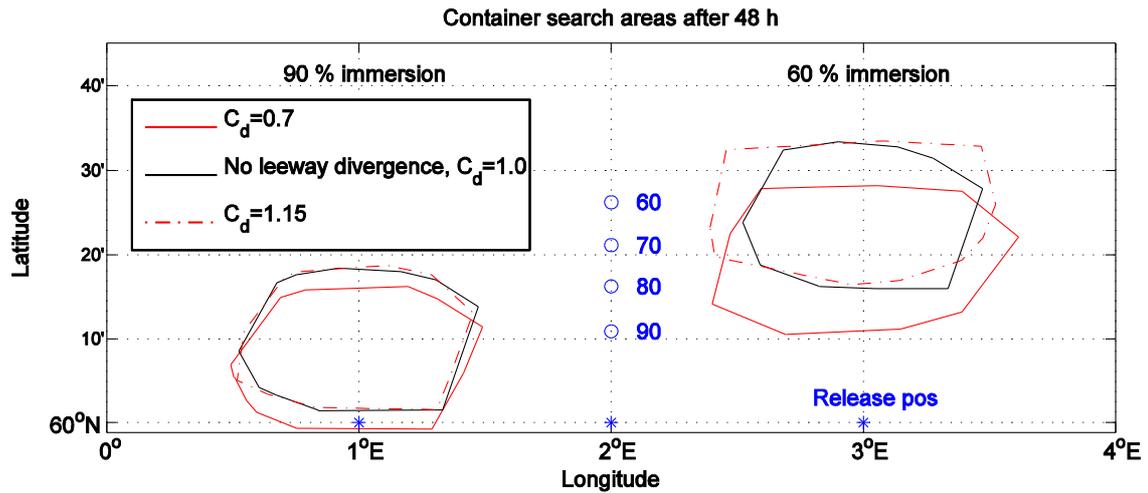

Figure 11. Container search areas for 90 % immersion (left) and 60 % immersion (right) after 48 hours. The displacements computed for immersion levels from 60 to 90 % with Eq (3) are shown in the middle for reference (blue rings marked 60 through 90) from a release position at 60º N, 002º E (release positions marked with "*"). The search areas were computed with the *Leeway* model (Breivik and Allen, 2008) with a constant southerly wind of 10 m s$^{-1}$ and no current. An ensemble of 500 particles was generated for each simulation. The convex hulls of all six simulations are shown with lower leeway coefficients corresponding to $C_d$=0.7 in red and high estimates corresponding to $C_d$=1.15 in red dashed. A crosswind leeway corresponding to a leeway divergence angle of 17º was used for these simulations. The black polygons correspond to estimates with a medium drag coefficient of $C_d$=1.0 and no leeway divergence.